\documentclass[11pt]{scrartcl}

\usepackage[utf8]{inputenc}
\usepackage[dvipsnames]{xcolor}
\usepackage[]{hyperref}
\hypersetup{
  colorlinks   = true, 
  urlcolor     = MidnightBlue, 
  linkcolor    = blue, 
  citecolor   = OliveGreen 
}

\usepackage[
	backend=biber,
	style=phys,
 	sorting=none,
]{biblatex}

\DeclareFieldFormat[inproceedings]{booktitle}{#1}
\DeclareFieldFormat*{title}{\mkbibitalic{#1}}
\DeclareFieldFormat[article]{volume}{\mkbibbold{#1}}
\DeclareFieldFormat[article]{journaltitle}{#1}
\renewbibmacro{in:}{} 
\AtEveryBibitem{\clearfield{issn}} 
\AtEveryCitekey{\clearfield{issn}}

\renewbibmacro*{volume+number+eid}{%
  \printfield{volume}%
  \setunit{\addcomma\space}%
  \printfield{number}%
  \setunit{\addcomma\space}%
  \printfield{eid} }

\AtEveryBibitem{%
  \ifboolexpr{not test {\ifentrytype{online}} }
    {\clearfield{month}%
     \clearfield{day}}
    {}%
}

\usepackage{amsmath,amssymb}
\usepackage[capitalize]{cleveref}

\usepackage[top=60pt,bottom=60pt,left=70pt,right=70pt]{geometry}
\usepackage{dsfont}
\newcommand{\id}{{\mathds{1}}}

\usepackage{qcircuit}

\usepackage{mathtools}

\DeclarePairedDelimiter\ket{\lvert}{\rangle}
\DeclarePairedDelimiterX\braket[2]{\langle}{\rangle}{#1 \delimsize\vert #2}

\newcommand{\Hi}{H_\text{Ising}}
\newcommand{\q}{q} 

\let\b b
\newcommand{\ms}{{\texttt{MS}}}

\begin{document}
\graphicspath{{./img/}}
\setlength{\abovedisplayskip}{13pt}
\setlength{\belowdisplayskip}{13pt}

\newcommand{\thetitle}{
Signal processing techniques for efficient compilation of controlled rotations in trapped ions
}
\title{ \Large \thetitle }

\author{ \bfseries \normalsize
Koen {Groenland}$^{1,2,*}$, Freek {Witteveen}$^{1,3}$, Kareljan {Schoutens}$^{1,2}$ and Rene {Gerritsma}$^{2}$}
\date{}
\maketitle

{ \vspace{-8mm} \scriptsize \noindent
{$^1$ QuSoft and Centrum Wiskunde en Informatica (CWI), Science Park 123, 1098 XG Amsterdam, the Netherlands} \\
{$^2$ Institute of Physics, University of Amsterdam, Science Park 904, 1098 XH Amsterdam, the Netherlands} \\
{$^3$ Korteweg-de Vries Institute for Mathematics, University of Amsterdam, Science Park 105-107, 1098 XG Amsterdam, the Netherlands} \\
{$*$ Contact: k.l.groenland@cwi.nl}
} \\

\begin{abstract}
{\normalsize
\noindent \textbf{Abstract:} Quantum logic gates with many control qubits are essential in many quantum algorithms, but remain challenging to perform in current experiments. Trapped ion quantum computers natively feature the M\o lmer-S\o rensen ({\ms}) entangling operation, which effectively applies an Ising interaction to all pairs of qubits at the same time. We consider a sequence of equal all-to-all {\ms} operations, interleaved with single-qubit gates that act only on one special qubit. Using a connection with quantum signal processing techniques, we find that it is possible to perform an arbitray SU(2) rotation on the special qubit if and only if all other qubits are in the state $\ket{1}$. Such controlled rotation gates with $N-1$ control qubits require $2N$ applications of the {\ms} gate, and can be mapped to a conventional Toffoli gate by demoting a single qubit to ancilla.
}
\end{abstract}

\vspace{6mm}
\section{Introduction}

Multiple-control gates arise in various quantum information processing tasks, ranging from basic arithmetic \cite{Saeedi2013} to Grover's search method \cite{Grover1996}, which famously requires a gate that performs an inversion around just a single specific state. The prototypical example is the \texttt{Toffoli}$_N$ gate, which performs a bitflip (Pauli-$X$) on the \emph{target} qubit if and only if $N-1$ \emph{control} qubits are in the state $\ket{1}$.

Albeit conceptually simple, gates with many control qubits remain challenging to implement in current experimental quantum computers. Because single-qubit gates are typically more accurate than entangling operations, an important challenge is to decompose multiqubit operations into circuits with as few as possible two-qubit gates.
To indicate, the \texttt{Toffoli}$_3$ requires at least 5 two-qubit gates \cite{Yu2013}, or when the \texttt{CNOT} gate is the only available entangling gate, then at least $6$ \texttt{CNOT}s are needed \cite{Shende2009}. For a larger number of qubits, the required number of basic operations or ancilla qubits grows steeply. 
The \texttt{Toffoli}$_N$ on $N$ qubits can be constructed through a circuit of \emph{depth} $O(\log(N))$, requiring $O(N)$ ancilla bits. With just a single ancilla, the best known circuits require $O(N)$ two-qubit gates \cite{Maslov2003,Maslov2016,He2017}. If no ancillas may be used, then a quadratic number of \texttt{CNOT}s are required \cite{Barenco1995}. The size of these circuits has been prohibitive in scaling up quantum algorithms on current quantum computer prototypes: even though various systems with a few tens of qubits have been reported, the largest multiple-controlled gate ever performed is, to our best knowledge, the \texttt{Toffoli}$_4$ \cite{Figgatt2017}.

A possible workaround is to replace operations by interactions between multiple qubits that natively arise in a quantum comptuter \cite{Wang2001,Isenhower2011, Groenland2018, Groenland2019a, Rasmussen2020}. 
Quantum computers based on trapped ions typically deal most naturally with the M\o lmer-S\o rensen gate ({\ms}) as basic entangling gate, which effectively applies an Ising interaction to all pairs of qubits for a specific amount of time \cite{Molmer:1999,Sackett:2000,Kim2009,Bruzewicz2019}. 
For a small number of qubits, fidelities of over $99\%$ have been reported \cite{Benhelm2008,Ballance:2016,Gaebler:2016}, and a large body of scientific work focuses on optimizing quantum circuits for this gate \cite{Zeng2005,Leibfried2018,Martinez2016,Maslov2017,Maslov2018,Figgatt2019}. To compare to the numbers of the \texttt{Toffoli}$_N$ gate, the best result we are aware of decomposes this operation into $3N-9$ {\ms} operations interleaved with single-qubit rotations \cite{Maslov2018}, requiring $\frac{N-2}{2}$ ancillary qubits and assuming that each operation may act on a different subset of qubits.

\paragraph{Overview of results}
In this work, we consider a gate set consisting of an all-to-all {\ms} gate that always acts on all $N$ qubits, together with arbitrary single-qubit rotations on just a single special qubit. Our main result is that this formalism allows one to perform rotations on the special qubit that depend on the state of the other qubits. We pay particular attention to the controlled rotation gate, which performs a single-qubit rotation on the special qubit if and only if all other qubits are in the state $\ket{1}$.

More precisely, we define the  M\o lmer-S\o rensen gate acting on all $N$ qubits in the system as
\begin{align}
\ms(\tau) = \exp \left( \frac{- i \tau }{4}  \sum_{j,k = 1}^N   X_j X_k  \right),
\label{eq:MSgate}
\end{align}
where ${X_j, Y_j, Z_j}$ denote the Pauli matrices acting on qubit $j$. In our case, $\tau$ is fixed for a given number of qubits $N$.
Moreover, we denote the conventional single-qubit rotations as
\begin{align}
R_x(\alpha) = \exp \left( -i X \frac{\alpha}{2} \right), \quad  R_y(\alpha) = \exp\left( -i Y \frac{\alpha}{2} \right), \quad R_z(\alpha) = \exp\left( -i Z \frac{\alpha}{2} \right).
\label{eq:rotations}
\end{align}
Then, the following circuit implements the operation $R_z(\alpha)$ on a target qubit when all other qubits (the \emph{controls}) are in the state $\ket{1}$:
%
\begin{align}
\raisebox{4.4em}{  
\Qcircuit @C=1 em @R=.8 em {
& &			&		&						& &		 	 &  	&					& \mbox{ Repeat for $j = L$ down to $1$ }	&				&				&	  &		&  \\
& & \ctrl{1} & \qw & \raisebox{-6.3 em}{ = }& & \gate{H} & \qw &\qw				& \multigate{3}{{\ms(\tau)}} 			& \qw 			&\qw				& \qw & \gate{H}  & \qw \\
& & \ctrl{1} & \qw & 					  	& & \gate{H} & \qw &\qw				& \ghost{{\ms(\tau)}} 	   				& \qw 			&\qw				& \qw & \gate{H}  & \qw \\
& & \ctrl{1} & \qw & 					  	& & \gate{H} & \qw &\qw				& \ghost{{\ms(\tau)}} 	   				& \qw 			&\qw				& \qw & \gate{H}  & \qw \\
& & \gate{ R_z(\alpha) } & \qw  & 			& & \qw	     & \qw &\gate{R_z(-\phi_j)}&\ghost{{\ms(\tau)}} 				& \gate{R_x(h)} &\gate{R_z(\phi_j)} & \qw & \gate{R_z(\phi_0)} & \qw
\gategroup{2}{9}{5}{12}{1.1em}{--}
\\
}
}
\label{eq:circuit}
\end{align}
Here, the top 3 lines represent $N-1$ control qubits, and a total of $L$ applications of the $\ms$ gate occur. $H = \frac{1}{\sqrt{2}}(X+Z)$ represents the Hadamard gate. The parameters $\tau, h$ are always $+\pi / N$ and $-\pi/N$ respectively, and the number of {\ms} pulses needed is, in this case, $L=2N$. The remaining unknown parameters, $\phi_j$, are discussed below. We call the resultant controlled rotation $C^{N-1} R_z(\alpha)$. Note that using a local basis transformation on the target qubit, this can be turned into any controlled-SU(2) operation. 
More generally, a similar circuit can implement any rotation $R_z(\alpha_q)$ on the target, where the rotation angle $\alpha_q$ depends on the \emph{number of control qubits in the state $\ket{1}$}. Such operations cost $L=4N$ applications of the {\ms} gate.

There is an important difference between $C^{N-1} R_x( \pm \pi )$ (a controlled $\mp i X$ rotation), and the conventional \texttt{Toffoli}$_N$ gate, because the factor of $i$ is \emph{not} a global phase. In \cref{eq:ConvToffoli}, we describe how the conventional \texttt{Toffoli}$_{N-1}$ can be retrieved by demoting one of the qubits to ancilla, following \cite{Rasmussen2020}.

The main body of this paper is devoted to linking the above circuit to the theory of equiangular composite gates, as introduced by Low, Yoder and Chuang \cite{Low2016}. This framework allows the construction of a single-qubit unitary operation $U(\theta)$ that depends in a complicated way on some parameter $\theta$, using a sequence of elementary gates consisting of $R_x(\theta)$ and arbitrary $\theta$-independent gates. The technique to efficiently calculate the appropriate gate sequence is often called signal processing. It proved useful in many recent breakthroughs in the design of quantum algorithms, such as quantum singular value transformations \cite{Gilyen2019stoc,Gilyen2019}, linear combinations of unitaries \cite{Childs2012}, and efficient Hamiltonian simulation \cite{Low2017,Low2019}.

In this paper, we use signal processing in a very different context. Our core result is that the action of the {\ms} gate can be interpreted as $R_x(\theta_q)$ on a special qubit, where $\theta_q$ depends on the number of ones among the other qubits. With this interpretation we can readily apply a known algorithm to efficiently compute the angles $\{ \phi_0, \phi_1, \ldots, \phi_L \}$ that implement the required gate, depending on $N$ and $\alpha$.

The theory and implementation of quantum signal processing is rather involved, and to lower the barrier to apply our results in practice, we provide Python code that calculates these parameters in Ref.~\cite{Groenland2019c}. In particular, the function \verb'crot_angles'( $N$, $\alpha$ ) returns precisely the list of angles $\phi_j$, which can be readily plugged into the circuit in \cref{eq:circuit}. Moreover, in \cref{sec:appendix}, we list an explicit circuit for $C^2 R_z(\pi)$ and tabulate values of the angles $\Phi$ for $N=3, 4, 5, 6$. This should allow anyone to use our results without the need to understand all the details of this paper.

\section{Composite gate techniques}
\label{sec:composite_gates}

This section reviews the theory of quantum signal processing and composite gates. We follow the notation from Low, Yoder and Chuang \cite{Low2016}, with some adjustments to make it consistent with the algorithm by Haah \cite{Haah2019} and our own notation of the {\ms} gate.

Let $\Phi = \{ \phi_0, \phi_1, \ldots, \phi_L \}$ be a list of angles (i.e.\@ real numbers that we interpret to be periodic with period $4\pi$)\footnote{Note that the rotations in \cref{eq:rotations} have periodicity $4 \pi$ and not $2 \pi$. To indicate, on input $\alpha=2 \pi$ each rotation becomes $-\id$. This makes a difference because we consider \emph{controlled} operations.} of length $L+1$. We can then define a single-qubit rotation of the form
\begin{align}
F^{\Phi}( \theta ) = R_z( \phi_0 ) \left[ \prod_{j=1}^{L} R_z( -\phi_j ) R_x( \theta ) R_z( \phi_j )\right].
\label{eq:composite_gate_from_rotations}
\end{align}
In other words, the function $F^\Phi$ maps an angle $\theta \in [0, 4 \pi)$ to an element of $SU(2)$, a unitary matrix with determinant $1$. Each of the $\theta$-dependent steps $ R_z( -\phi_j ) R_x( \theta ) R_z( \phi_j ) $ corresponds to a rotation around a vector pointing along the equator of the Bloch sphere, where this vector is given by $\cos( \phi_j /2 ) X - \sin( \phi_j /2 ) Y$.

Any such matrix $F^\Phi(\theta)$ can also be decomposed in a basis of Pauli matrices%
\footnote{Note that there exist two different conventions for expanding $F^{\Phi}(\theta)$ into Pauli matrices: we follow the notation by Haah \cite{Haah2019}, whereas Low et al. work with slightly different labeling $ABCD \rightarrow ACDB$ \cite{Low2016,Low2017,Low2019}. This is equivalent to a change of basis on the Pauli matrices.}%
,
\begin{align}
F^{\Phi}(\theta) = A(\theta) \id + i B(\theta) X + i C(\theta) Y + i D(\theta) Z.
\end{align}
Here, $A, B, C, D$ are real functions of $\theta$. They have to be $4 \pi$ periodic in $\theta$, and by unitarity, they satisfy the normalization condition $A(\theta)^2 + B(\theta)^2 + C(\theta)^2 + D(\theta)^2 = 1$. Moreover, it was found that $A$ and $D$ are \emph{symmetric} functions of $\theta$, whilst $B, C$ are \emph{anti-symmetric} in $\theta$ (i.e.\@ $A(\theta) = + A(-\theta)$ and $C(\theta) = - C(-\theta)$) \cite{Haah2019}. 

With the above properties, it is convenient to express  $A, B, C, D$ in their cosine or sine series. As a technical detail, an expansion over $4\pi$ periodic function takes the form $A = \sum_{k=0}^M \tilde{a}_k \cos(k \theta / 2)$, but due to the structure of SU(2), only frequencies of $k$ odd (when $L$ is odd) or $k$ even (when $L$ is even) can have nonzero coefficients $\tilde{a}_k$ (and likewise for $B,C,D$)%
\footnote{To indicate, $R_x(\theta + 2 \pi) = - R_x(\theta)$. Therefore, for sequences with an \emph{even} number of applications of $R_x(\theta)$, the functions $A,B,C,D$ are symmetric under a $2\pi$ shift in $\theta$.  An odd number of applications require basis functions to be anti-symmetric under a $2\pi$ shift.}.
We choose to work only with the simpler case of $L$ even, where all functions become $2\pi$ periodic in $\theta$:
\begin{align}
A(\theta) = \sum_{k=0}^{L/2} a_k \cos(k \theta), \hspace{1cm}
B(\theta) = \sum_{k=1}^{L/2} b_k \sin(k \theta) \label{eq:cosineseries} 	\\
C(\theta) = \sum_{k=1}^{L/2} c_k \sin(k \theta), \hspace{1cm}
D(\theta) = \sum_{k=0}^{L/2} d_k \cos(k \theta) \nonumber
\end{align}
In this notation, the degree of the series $L/2$ turns out to be half of the number of $\theta$-dependent steps in \cref{eq:composite_gate_from_rotations} \cite{Low2016}.

The reverse is also true: for any set of series $A,B,C,D$ that is properly normalized and has largest degree $L/2$, there exists a sequence of angles $\Phi$ of length $L+1$ such that \cref{eq:composite_gate_from_rotations} holds. Retrieving these angles $\Phi$ from known series $A,B,C,D$ can be done efficiently on a classical computer using the algorithm by Haah \cite{Haah2019}, which is implemented in Python in Ref.~\cite{Roberts2019}. During the completion of this manuscript, two alternative approaches to retrieve $\Phi$ were announced \cite{Ding2020,Chao2020}. 

This backwards-engineering step is precisely what we will exploit. In \cref{sec:ms} we indicate how {\ms} pulses can be interpreted as a single-qubit rotation of the form $R_x(\theta_q)$ on the target qubit, where $\theta_q$ depends on the state of the control qubits. Then, in \cref{sec:full_composite_gate}, we find achievable cosine and sine series $A,B,C,D$ that correspond precisely to the gate we aim to obtain, with degrees as low as possible. In fact, we only have to supply the algorithm with $A$ and $B$, as it determines a suitable $C$ and $D$ by itself using the normalization and even/oddness.

\paragraph{} The remainder of this section consists of technical notes that one might want to skip on a first read.
Firstly, we note that Haah's algorithm is slightly more general, allowing $A$ to be chosen either even or odd, and similar (independently) for $B$. The downside is that in such cases, the $\theta$-dependent single-qubit gates $R_z( -\phi_j ) R_x( \theta ) R_z( \phi_j )$ have to be replaced by $\theta$-rotations around an arbitrary vector on the Bloch sphere. Because we did not find more efficient results in this formalism, we choose to stick with the simpler notation introduced above. Still, future work might exploit these extensions to improve on our results.

We will later see that, to obtain the advertised $C^{N-1} R_z(\alpha)$ gate, it is essential that we have control over the function $D$, which does not directly serve as input. However, we will ensure that $D$ is non-zero only at the special points $\theta=0$ or $\theta=\pi$, where the anti-symmetric functions $B,C$ are always $0$. Exploiting $D^2 = 1-A^2$ at these points, we can produce $R_z(\alpha)$ by only pinning $A$ to a specific value.

\section{M\o lmer-S\o rensen in the context of single-qubit composite gates}
\label{sec:ms}
In this section, we argue how the rotation $R_x(\theta)$ (acting on a single qubit, but having $\theta$ as free parameter) can be related to the {\ms} operation (acting on $N$ qubits). We will see that a circuit of the form \cref{eq:circuit} can be broken down into invariant two-dimensional subspaces. The operation on each subspace can be described by $F^{\Phi}(\theta_q)$, where $\theta_q$ depends on the state of the control qubits.

We consider a set of $N$ qubits, which we label by $[N] = \{ 0, 1, \ldots, N-1 \}$. The $\ms(\tau)$ operation is interpreted as the evolution of a certain Ising model under Schr\"{o}dinger's equation for a time $\tau$. For now, let us analyse the Ising model in the $Z$-basis rather than the $X$-basis, and allow for arbitrary interaction strength $w_{jk}$ between any pair $j,k \in [N]$ of qubits. The corresponding Hamiltonian reads
\begin{align}
\Hi = \frac{J}{2} \sum_{ \substack{j,k \in [N] \\ j < k}} w_{jk}  Z_j Z_k.
\label{eq:Hising}
\end{align}
The variable $J$ sets an energy scale. Interactions of this form have been experimentally observed between $N=53$ atoms in a Paul trap \cite{Zhang2017a}.
In our case, we refer to qubit $0$ as our special \emph{target} qubit, and the other qubits will be called \emph{control} qubits. We denote our quantum states in the computational basis as $\ket{\b_0, \vec{\b}}$, where $\b_0 \in \{0,1\}$ represents the state of the target, and $\vec{\b} \in \{0,1\}^{N-1}$ denotes the states of the control qubits. The states $\ket{\b_0, \vec{\b}}$ are the eigenstates of $\Hi$, whose energies we denote by $E_{\b_0, \vec{\b}}$.

When we include arbitrary rotations on the target qubit, the Hilbert space decomposes into conserved subspaces, each of which can be labeled by the state $\vec{\b}$ of the control qubits:
\begin{align}
\mathcal{H}_{\vec{\b}} = \text{span}( \ket{0, \vec{\b}},  \ket{1, \vec{\b} } ).
\label{eq:Hdecompose}
\end{align}
Within each of these subspaces, the Ising Hamiltonian $\Hi$ acts as
\begin{align}
H_{\vec{\b}} &= \begin{pmatrix}
E_{0,\vec{\b}} 	& 0 \\
0 				& E_{1,\vec{\b}}
\end{pmatrix} \nonumber \\
&=
\begin{pmatrix}
\Delta_{\vec{\b}} / 2 	& 0 \\
0						& -\Delta_{\vec{\b}} / 2
\end{pmatrix}
+ \bar{E}_{\vec{\b}} \ \mathds{1},
\label{eq:Hx}
\end{align}
where we defined the energy gap $\Delta_{\vec{\b}} = E_{0,\vec{\b}} - E_{1,\vec{\b}}$ and the mean energy $\bar{E}_{\vec{\b}} = \frac{E_{0,\vec{\b}} + E_{1,\vec{\b}}}{2}$.

\begin{figure}
\centering
\includegraphics[width=.24\linewidth]{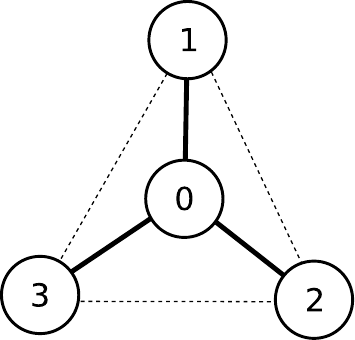}
\hspace{1cm}
\includegraphics[width=.43\linewidth]{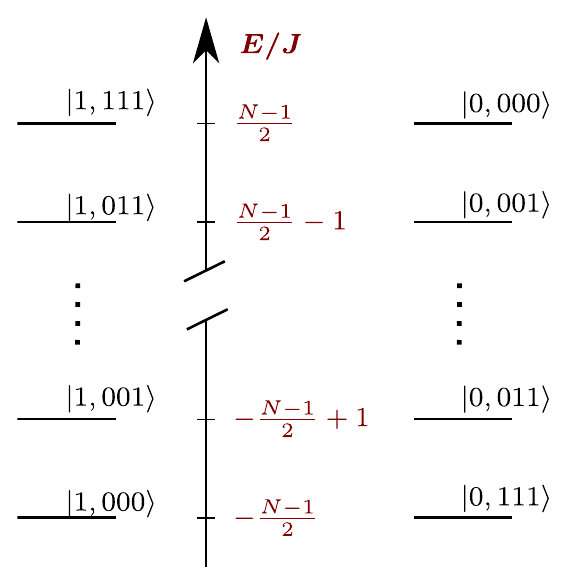}
\caption{The left image sketches the star connectivity for a system of size $n=3$. The right-hand side displays the energy spectrum of $\Hi$ of such a system, in units of $E/J$. }
\label{fig:spectrum}
\end{figure}

Now, consider the special configuration of couplings
\begin{align}
 w_{0,k} &= 1 \quad \forall \ k  \in \{ 1, \ldots, N-1 \}   && \text{(Star couplings)}  \label{eq:star} \\
 w_{j,k} &= 0 \quad \forall \ j \geq 1. \nonumber
\end{align}
This gives rise to a star-shaped connectivity, where all control qubits are coupled to the target, but not among each other. Now, $\Hi$ has a spectrum as indicated in Fig. \ref{fig:spectrum}: the ground energy is $E=- J (N-1)/2$ when all control qubits are \emph{different} from the target, and $+J$ energy is added for each control qubit that is the same. The complete eigensystem is
\begin{align*}
\{ \ket{0, \vec{\b} } : |\vec{\b}| = \q \}  & & E_{0,\vec{\b}} = J \left( \frac{N-1}{2}-q \right) && \text{(Star couplings)}  \\
\{ \ket{1, \vec{\b} } : |\vec{\b}| = \q \} & & E_{1,\vec{\b}} = -J \left( \frac{N-1}{2}-q \right)
\end{align*}
where $|\vec{\b}| = \q \in \{ 0, \ldots, N-1 \}$ denotes the Hamming weight (i.e.\@ the number of ones) of the bitstring $\vec{\b} \in \{ 0, 1 \}^{N-1}$. For this configuration, $H_{\vec{\b}}$ takes a particularly simple form, as the gap $\Delta_{\vec{\b}}$ is given by
\begin{align}
\Delta_{\vec{\b}} = J (N-1-2 \q),
\label{eq:energygap}
\end{align}
hence is the same for all subspaces corresponding to the same Hamming weight $\q$. Moreover, the mean energy vanishes for all subspaces, i.e.\@ $\bar{E}_{\vec{\b}} = 0$.

Now, let us turn on the couplings $w_{jk}$ between the control qubits, leaving the couplings connected to the target fixed at $w_{0,k} = 1$. The energy difference compared to the star coupling case can now only depend on the state $\vec{\b}$ of the controls, not on the state of the target. Hence, the energy gap $\Delta_{\vec{\b}} = E_{0,\vec{\b}} - E_{1,\vec{\b}}$ will not change as a response to this. In fact, the only parameter of $H_{\vec{\b}}$ that changes is $\bar{E}_{\vec{\b}}$, leading to a subspace-dependent overall shift in energy.

\paragraph{}
The link between $\Hi$ and the {\ms} gate is as follows. Choosing uniform all-to-all couplings ($w_{jk} = 1$) and unitarily evolving for a time $t = \tau / J$ precisely implements the $\ms(\tau)$ gate, up to a rotation between the $X$ and $Z$ eigenbases:
\begin{align*}
 \ms(\tau) = H^{\otimes N} \cdot  \exp( -i \Hi \tau / J ) \cdot  H^{\otimes N}
\end{align*}
Here, $H$ denotes the Hadamard gate. Within the conserved subspace $\mathcal{H}_{\vec{\b}}$, up to an overall phase, this operation can be written as
\begin{align*}
\ms(\tau) \sim R_x(   \Delta_{\vec{\b}} \ \tau / J  ) \ e^{ -i \bar{E}_{\vec{\b}} \tau / J }
\end{align*}
For later convenience, after\footnote{Because $\ms(\tau)$ and $R_x(h)$ commute, they can be performed in any order, or even simultaneously.} each {\ms} gate we allow a single-qubit rotation $R_x( h )$ to be applied on the target qubit. The combined rotation takes the form
\begin{align*}
\ms(\tau) R_x(h) \sim R_x( \theta_q ) \ e^{ -i \bar{E}_{\vec{\b}} \tau / J }  \quad \text{ where } \quad \theta_q =  \frac{\Delta_{\vec{\b}} \ \tau}{J} + h = (N-1 - 2 q) \tau + h.
\end{align*}
We collect the rotation angles $\theta_q$ using the set notation $\Theta = \{ \theta_q \}_{q=0}^{N-1}$, representing all the rotation angles $\theta_q$ that can occur due to a combination of ${\ms}(\tau)$ and $R_x(h)$. The free parameters $\tau, h$ allow us to spread and shift these relevant angles over the unit circle.

This approach connects the circuit in \cref{eq:circuit} with the composite gate defined in \cref{eq:composite_gate_from_rotations}: the function $F^\Phi(\theta)$ can be implemented by a sequence of {\ms} gates (that effectively perform $R_x(\theta)$) interleaved with $R_z(\phi_j)$ rotations. The angles $\theta$ that can occur are precisely those in $\Theta$, which depend on the state of the control qubits. Pinning the rotation $F^\Phi(\theta_q)$ that takes place on the \emph{target} qubit whenever the control qubits have Hamming weight $q$ is equivalent to pinning the values $A(\theta_q), B(\theta_q), C(\theta_q), D(\theta_q)$.

A last detail is the subspace-dependent phase due to $\bar{E}_{\vec{\b}}$, which can be straightforwardly tracked in the case of equal all-to-all interaction ($w_{jk} = 1 \ \ \forall j, k$), as is the case in our definition of the {\ms} gate. The permutation symmetry among qubits makes it straightforward to calculate the $N$ unique values of $\bar{E}_{\vec{\b}}$. However, in the context of the circuit in \cref{eq:circuit}, we follow a more intuitive approach. We observe that all the terms $X_j X_k$ in the {\ms} gate commute. Moreover, for the \emph{control} qubits, each subsequent $X_j X_k$ pulse is not interleaved with single-qubit gates, hence rotation angles are additive. Each of these terms associates a phase to a pair of qubits $(j,k)$. When $\tau$ and $L$ are chosen such that $\tau L = 2\pi$, each of these phases reset regardless of the qubit state. Thus, with $\tau = \pi / N$, any choice of $L$ that is a multiple of $2N$ guarantees that subspace-dependent phases $\bar{E}_{\vec{\b}}$ reset in our circuit.

\section{The full composite gate}
\label{sec:full_composite_gate}

\begin{figure}
\centering
\includegraphics[width=.4\linewidth]{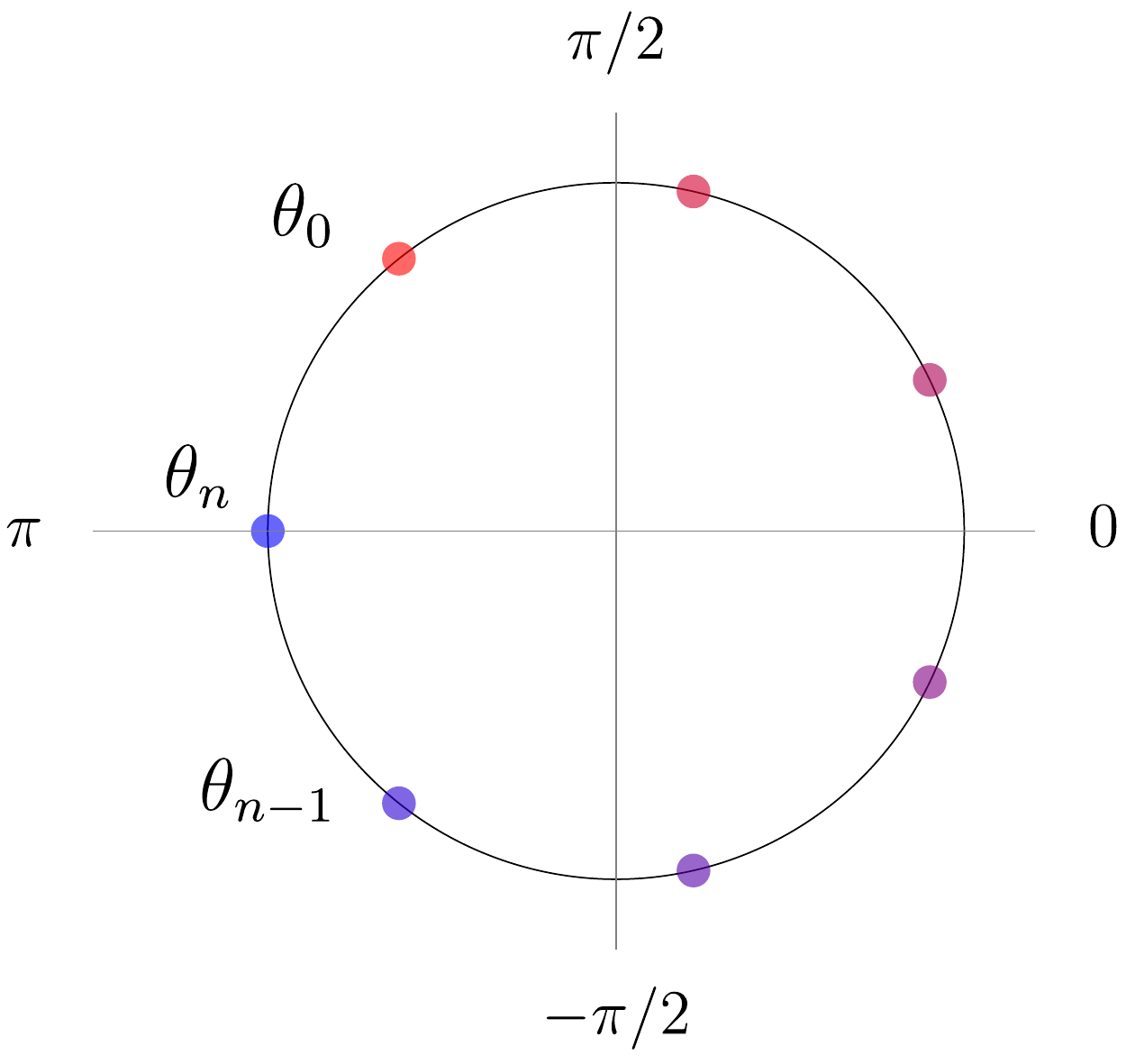}
\includegraphics[width=.55\linewidth]{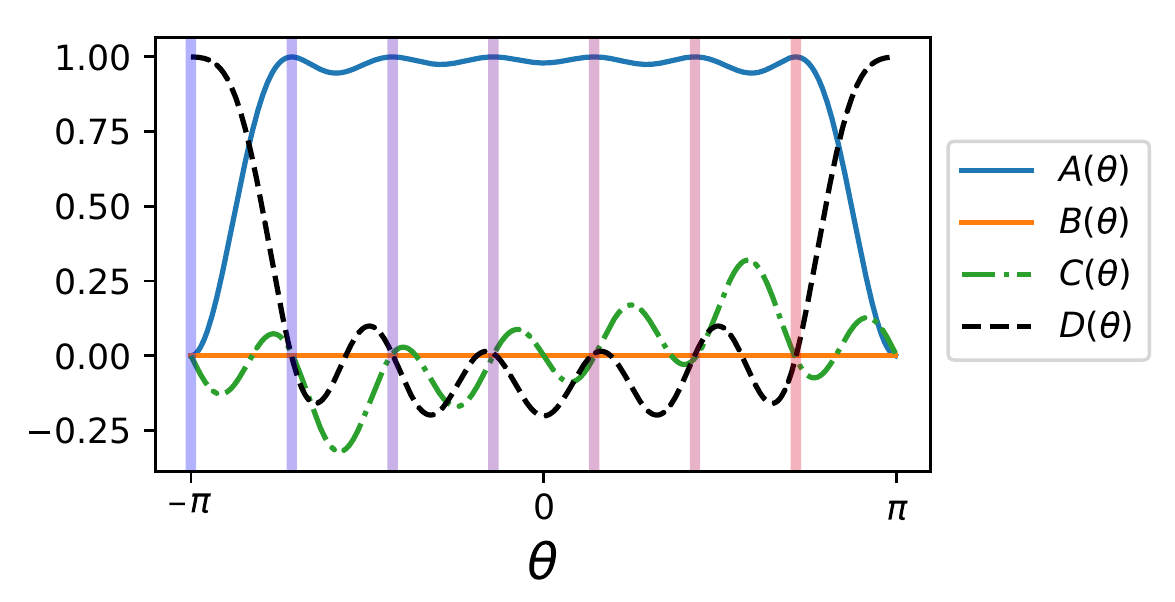}
\caption{An indication of the composite gate parameters for the case $N=7$ and $\alpha = \pi$. Left: Our choice of $\tau, h$ spreads the values $\theta_q$ uniformly over the unit circle, with the special point $\theta_n = -\pi$. Right: The cosine and sine series $A,B,C,D$ which were obtained using our fitting strategy. }
\label{fig:ABCD}
\end{figure}

Let us now consider the functions $A,B,C,D$ that implement a controlled rotation. The gate $C^{N-1} R_z(\alpha)$ corresponds to the choice
\begin{align}
 \text{ if } q \neq N-1: \quad & A(\theta_q) = 1,   \label{eq:ADvalues} \\
 \text{ if } q = N-1:  	\quad  & A(\theta_q) = \cos(\alpha/2), \quad D(\theta_q) = \sin(\alpha / 2). \nonumber
\end{align}
We first choose convenient values values for $\tau$ and $h$. We aim to have all relevant angles $\theta_q$ spread out in the interval $[0,2\pi]$ as much as possible, as in \cref{fig:ABCD}, such as to obtain series with the lowest possible degree and to allow more leeway in experimental control. Moreover, we choose $\theta_{q=N-1} = \pi$ to be at a symmetric point of the even functions $A$ and $D$, such that the function value $D(\theta_{q=N-1})$ is never repeated within the function's period.  Therefore, we choose\footnote{Note that other choices, such as setting $\theta_{q=N-1} = 0$ would work equally well.}
\begin{align*}
\theta_q = \pi -  \frac{ 2 \pi }{N} (q+1).
\end{align*}
This set of angles is obtained by setting
\begin{align*}
h = - \frac{\pi}{N}, \quad \tau = \frac{\pi}{N}.
\end{align*}

\paragraph{Fitting $A$ and $B$}
Our next step is to find low-degree cosine/sine series for $A$ and $B$ that satisfy \cref{eq:ADvalues}. We choose to work only with even functions and set $B=0$. The coefficients of $a_k$ as in \cref{eq:cosineseries} can be solved using the $N$ constraints on $A$ given in \cref{eq:ADvalues}. In fact, due to symmetry we need only enforce these values for\footnote{We use $\lfloor a \rfloor$ to denote that $a$ is rounded down to the nearest integer. Similarly, $\lceil a \rceil$ denotes rounding up.} $q = \lfloor \frac{N}{2} \rfloor$ up to $q=N-1$. 

Unfortunately, such fits would generally not respect the normalization condition $A^2 + B^2 + C^2 + D^2 = 1$. To enforce $|A(\theta)| \leq 1$, we restrict the derivatives $A'(\theta_q) = 0$. This way, whenever $A(\theta_q)=1$, this corresponds to a maximum of the function $A$. An additional advantage is that slight over- or under-rotations of $\theta$ due to an inaccurate {\ms} gate do not affect the resulting operation, to first order in $\theta$. Exploiting again the symmetries of $A$, the only values of $q$ for which these derivatives are relevant are $\lceil \frac{N}{2} \rceil$ up to $N-2$, because the cosine series already has zero derivative at the points $\theta = 0, \pi$. All in all, we have $N$ constraints on the function $A(\theta)$, hence we can parametrize it as a series of degree $N-1$ (because the series starts counting at $k=0$).

With this choice of $A(\theta)$, any choice of $B,C,D$ that respects the normalization condition is automatically a valid choice for the $C^{N-1}R_z(\alpha)$ gate. Firstly, at the points $\theta_q$ for $q \neq N-1$, we have set $A = 1$ and hence at those points, $B=C=D=0$. Moreover, the point $\theta_{N-1} = \pi$ is a zero for any sine series, so normalization requires $|D(\theta_{N-1})| = \sin( \alpha/2 )$.\footnote{In our case, where we rely on Haah's algorithm to choose the precise form of $D(\theta)$, we may end up with the `negative' branch where $D(\theta_{N-1}) = - \sin(\alpha/2)$. In general, it is straightforward to map our circuit into its inverse: the circuit in \cref{eq:circuit} should be ran backwards, and the $X$ and $XX$ operations should be conjugated by setting $\phi_j \rightarrow \phi_j + \pi$. } The behavior of $B,C,D$ at points other than $\theta_q$ is irrelevant for our purposes.

\paragraph{From cosine series to composite gate}
Having found the series representation of $A$ and $B$, we can input these in the algorithm by Haah \cite{Haah2019}. It finds the angles $\Phi$ that allow us to form the circuit in \cref{eq:circuit} with $L= 2 N - 2$.

\paragraph{On relative phases due to $\bar{E}_{\vec{\b}}$}
In its current form, the circuit in \cref{eq:circuit} that uses a sequence of length $L = 2N-2$ will also introduce subspace-dependent phases due to $\bar{E}_{\vec{\b}}$. The simplest solution is to round $L$ up to $L=2N$ by adding two addition steps with $\phi_{2N-1} = 0$ and $\phi_{2N} = \pi$. These two extra steps have no net effect on the target qubit, but make sure that the {\ms} gate performs a full $2\pi$ rotation on all pairs of control qubits.

\paragraph{Turning the $C^{N-1}R_z(\alpha)$ into a conventional \texttt{Toffoli}}
The rotation of the form $C^{N-1}R_z(\pi)$ fundamentally differs from a conventional \texttt{Toffoli}$_{N}$ or controlled-$Z$ operation due to the additional phase $i$ that appears if and only if the target qubit is rotated. To get rid of this extra factor in the subspace where $q=N-1$, we propose the following circuit, as inspired by Ref.~\cite{Rasmussen2020}:
\begin{align}
\raisebox{3.5em}{
\Qcircuit @C=1 em @R=1 em {
& \ctrl{1}  &  \qw   & \raisebox{-6.3 em}{=} 	& & \qw			 	 & \qw 	& \ctrl{1}       	& \qw &  \qw & \qw 		\\
& \ctrl{1}  &  \qw   & 					 	& & \qw			 	 & \qw 	& \ctrl{1}       	& \qw &  \qw & \qw 		 \\
& \ctrl{1}  &  \qw   &  						& & \qw			 	 & \qw 	& \ctrl{1}       	& \qw &  \qw & \qw 		 \\
& \ctrl{1}  &  \qw   &  						& & \qw			 	 & \qw 	& \ctrl{1}       	& \qw &  \qw & \qw 		\\
& \targ 	&  \qw   &                      	& & \qw	 	 & \gate{H}  	& \ctrl{1}   		& \gate{H} &  \qw & \qw   \\
&         	&  	  	& 						& &  & \lstick{\ket{0}}  & \gate{R_z(2 \pi)} & \qw &   \lstick{{\ket{0}}}
 } 	
} 	
\label{eq:ConvToffoli}
\end{align}
The philosophy here is that the $C^{N-1} R_z(2\pi)$ gate applies the operation $-\mathds{1}$ if all control qubits are $\ket{1}$ and the operation $\mathds{1}$ otherwise. Hence, this is the same as a controlled-$Z$ gate acting only on the control qubits. A conjugation by the Hadamard $H$ on a single qubit maps this into a conventional \texttt{Toffoli}$_{N-1}$, with the respective qubit taking the role of target.

\section{Discussion and possible extensions}

The number of Ising pulses $L=2N$ needed to implement the $C^{N-1} R_z(\alpha)$ is very modest and does not depend on $\alpha$. Still, for small $N$, there exist highly optimized decompositions that are more optimal than our results. For example, Refs. \cite{Martinez2016,Maslov2018} find decompositions of, respectively, the $N=3$ and $N=4$ \texttt{Toffoli} gates into a mere 3 ${\ms(\tau)}$ pulses (albeit with $\tau$ not equal for all pulses). Moreover, the $C^2 R_x(\pi)$ operation can be implemented in just 4 \texttt{CNOT} operations%
\footnote{The circuit can be obtained from the textbook \cite{Nielsen2010}, page 182, Fig. 4.9, when discarding the last six gates on the top two qubits. Thanks to an anonymous referee for pointing this out to us.}. 

Our results are of main interest thanks to the favourable scaling with increasing $N$. The best similar result we are aware of is Ref.~\cite{Maslov2018}. It describes circuits for the \texttt{Toffoli}$_N$ gate that use $3N-9$ {\ms} gates and $\frac{N-2}{2}$ ancillae for even $N$, and $3N - 6$ {\ms} gates and $\frac{N-1}{2}$ ancillae when $N$ is odd, both of which were major improvements over previously known results. Each of these {\ms} gates has to act on varying subsets of qubits, and the rotation angle $\tau$ can differ per step. Using the circuit in \cref{eq:ConvToffoli}, our proposal implements the \texttt{Toffoli}$_N$ using $2(N+1)$ {\ms} gates, significantly improving the scaling with $N$ and requiring a mere $1$ ancilla. Moreover, our proposal has the added advantage that all {\ms} gates are equal: they act equally on all pairs of qubits, and require the same rotation angle $\tau$ in each gate. This simplicity is an experimental advantage, putting less constraints on control hardware and requiring fewer unique {\ms} pulses to be optimized \cite{Schindler2013}.

\paragraph{}
Our proposal is particularly suitable for the trapped ion quantum information processor described in Ref.~\cite{Schindler2013}. Here, the {\ms} interactions are equal between all ions since they are mediated by longitudinal center-of-mass motion in the linear ion crystal. Furthermore, single ion addressing is possible with a focussed laser beam  allowing the implementation of the single-qubit rotations. In this experimental setup, high fidelity {\ms} gates have been implemented with up to 14 ions \cite{Monz2011,Erhard2019}. Our scheme remains viable in future array-based trapped ion computers based on ion shuttling and crystal merging \cite{Kielpinski:2002}. In such a system, {\ms} gates can remain confined to small sub-crystals.

Alternatively, much longer crystals containing $>50$ ions could be considered such as in Ref.~\cite{Zhang2017a}. However, in these systems the interactions are mediated by \emph{transverse} phonons, resulting in couplings of the form   $w_{jk}\sim |j-k|^{-\alpha}$ with $0 \leq \alpha \leq 3$ in theory, but $0.5 \leq \alpha \leq 1.8$ in practice. 

In general, a non-uniform Ising model would lead to two types of issues. Firstly, we consider non-uniform couplings $w_{jk}$ only between the \emph{control} qubits. The resultant operation would look like a normal controlled rotation, perturbed by additional $XX$ interactions among the controls. These phases can in principle be undone by a single Ising evolution that affects \emph{only} the controls and has connection strengths $-w_{jk}$, i.e. the negative values of the previous operation. However, there is no straightforward method of inverting the sign of the interactions for interactions mediated by transverse phonons \cite{Kim2009}. Alternatively, one may keep applying non-uniform operations with weights $w_{jk}$ until all the relative phases reset. The problem then reduces to finding multiples of $w_{jk}$ which are all integer multiples of some base strength. It seems likely that newly developed techniques for finding robust gate operations based on multiple laser frequencies could be successfully applied to find such interaction matrices \cite{Leung:2018,Shapira:2019}. 

A second type of issue arises when the connection strengths $w_{0k}$ to the \emph{target} qubit are no longer equal, which causes a larger set of possible values $\theta_q$. This results in series  $A,B,C,D$ with a larger degree, and hence sequences of more MS pulses. 

We conclude that our proposal should be within experimental reach on devices that rely on longitudinal phonon interactions, whereas for transversal interactions further optimizations are needed.

\paragraph{}
From a theoretical perspective, many further optimizations and extensions of our protocol should be possible. For example, much can be learned from combining the intuition in Ref.~\cite{Maslov2018} and our results. In particular, making $\tau, h$ different per step, and adding single-qubit rotations on the control qubits, might greatly extend the possibilities of our framework. Moreover, in the large-$N$ limit, one might consider functions $A,B,C,D$ that merely \emph{approximately} implement a required gate, in exchange for a reduction of the number of steps $L$.

From a computer science perspective, we would be very interested to see lower bounds on the number of {\ms} gates needed for certain operations. Moreover, it seems unclear to us how the gates \texttt{Toffoli}$_N$ and $C^{N-1} R_z(\alpha)$ compare when used in algorithms in practice, and we hope that future research will find how many of each are required in realistic cases. Some results in this direction can be found in Ref. \cite{Maslov2016}.

\paragraph{}
As another possible extension of our protocol, assuming again uniform weights $w_{jk}=1$, one can make controlled rotation gates that depend in a more general way on the \emph{Hamming weight} of the controls. That is, one could try to implement different functions $A(\theta), B(\theta), C(\theta), D(\theta)$ that act differently for each $\theta_q \in \Theta$ corresponding to a specific Hamming weight. For example, it should be possible to make a gate that performs $R_x(\alpha_0)$ on the target whenever all controls are $0$, and $R_x(\alpha_1)$ whenever the controls have weight $q=1$, etc, by pinning the values $A(\theta_q) = \cos( \alpha_q )$ and $B(\theta_q) = \sin( \alpha_q )$. More generally, one could try to solve for any weight-dependent operation mixing $X$, $Y$ and $Z$ operators, under the constraint that the functions $A,B,C,D$ are properly normalized. This can be a challenging problem and is very similar to filter design problems in discrete time signal processing \cite{Oppenheim1999} and as also suggested by \cite{Low2016}, one can take advantage of the wealth of existing techniques from this domain to search appropriate solutions. 

Lastly, we note that our results may also be of interest for the engineering of related many-body interactions in trapped ion quantum simulators \cite{Barreiro2011,Lanyon2011} and may be a starting point for implementing more general many-body interactions decomposed out of native quantum gates \cite{Mueller2011,Weimer2010}.

\section{Conclusion}
We make a connection between the all-to-all M\o lmer-S\o rensen gate and signal processing techniques. Our main result is that controlled rotations of the form $C^{N-1} R_z(\alpha)$ can be formed by a circuit consisting of $2N$ {\ms} gates plus single-qubit gates that act only on the target qubit. This operation is easily mapped to a \texttt{Toffoli}$_{N-1}$ gate. Each of the required building blocks has been performed at high fidelity in recent experiments, indicating that our formalism can be realistically applied on near-term quantum computers. We also identify various extensions that may lead to further improvements and generalizations of our protocol.

\section*{Acknowledgments}
K. Groenland and K. Schoutens are supported by the QM\&QI grant of the University of Amsterdam, supporting QuSoft. R. Gerritsma is supported by the Netherlands Organization for Scientific Research (Grant 680.91.120).

\printbibliography


\appendix
\clearpage
\section{An example circuit}
\label{sec:appendix}
\begin{figure}[h]
\centering
\includegraphics[width=\linewidth]{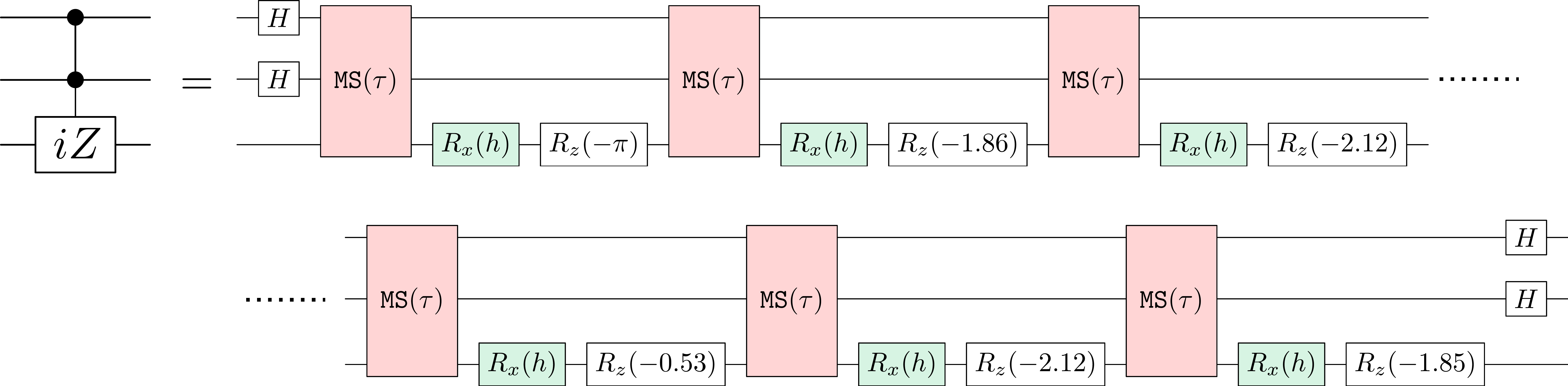}
\caption{The circuit for the $C^2 R_z(-\pi)$ operation derived using the method presented in this paper. Note that adjacent $R_z(\phi_j)$ operations have been merged, leading to a more compact form than \cref{eq:circuit}. Generalization to larger a larger number of qubits $N$ is straightforward using the numbers in \cref{tab:valuesPhi}. }
\label{fig:example_circuit}
\end{figure}

\cref{fig:example_circuit} shows an explicit circuit for the operation  $C^{2} R_z(-\pi)$, i.e. the double-controlled $iZ$ gate. Here, we combined consecutive rotations of the form $R_z(\phi_j) R_z(-\phi_{j-1})$ into a new rotation $R_z( \widetilde{\phi}_j )$ to shorten the circuit. For other system sizes $N=3,4,5,6$, the related angles $\widetilde{\phi}_0, \ldots, \widetilde{\phi}_L$ for the $C^{N-1}R_z(-\pi)$ operation can be foud in \cref{tab:valuesPhi}, where again the tilde indicates that adjacent rotations with angles $\phi_j, -\phi_{j-1}$ have been combined. 

\vspace{2cm}

\vspace{2cm}

\setlength{\tabcolsep}{4pt}
\renewcommand{\arraystretch}{1.5}

\begin{table}[h]
\centering
{ \footnotesize 
\begin{tabular}{c|c|ccccccccccccc}
    	& $\tau = -h $    & $\widetilde{\phi}_0$ & $\widetilde{\phi}_1$ &  $\widetilde{\phi}_2$ &  $\widetilde{\phi}_3$ &  $\widetilde{\phi}_4$ &  $\widetilde{\phi}_5$ &  $\widetilde{\phi}_6$ &  $\widetilde{\phi}_7$ &  $\widetilde{\phi}_8$ & $\widetilde{\phi}_9$ & $\widetilde{\phi}_{10}$ & $\widetilde{\phi}_{11}$ \\
    	\hline
$N=3$ & $\pi / 3$ &
-1.855 & -2.118 & -0.525 & -2.118 & -1.855 & $-\pi$ & $0$
\\
$N=4$ & $\pi/4$   &
-2.366 & -1.564 & 1.577 & 1.55 & 1.577 & -1.564 & -2.366 &
 $-\pi$ & $0$
\\
$N=5$ & $\pi/5$   &
-2.61 & -1.098 & 1.417 & -1.116 & -2.041 & -1.116 & 1.417 & -1.098 & -2.61 &
$-\pi$ & $0$
\\
$N=6$ & $\pi/6$   &
-2.745 & -0.79 & 1.146 & -1.155 & 0.81 & 2.312 & 0.81 & -1.155 & 1.146 & -0.79 & -2.745
& $-\pi$ 
\end{tabular}
}
\caption{The rotation angles $\widetilde{\phi}_j$ that implement a controlled-$iZ$ operation with $N-1$ controls, in the way indicated \cref{fig:example_circuit}. Note that we chose the angle numbered $2N$ to be zero, such that it can be omitted. }
\label{tab:valuesPhi}
\end{table}

\end{document}